\documentclass{article}

\usepackage{dsfont}
\usepackage{newunicodechar} 

\usepackage{arxiv}

\usepackage[utf8]{inputenc} 
\usepackage[T1]{fontenc}    
\usepackage[colorlinks, citecolor=green, linkcolor=red]{hyperref}      
\usepackage{url}            
\usepackage{booktabs}       
\usepackage{amsfonts}       
\usepackage{nicefrac}       
\usepackage{microtype}      
\usepackage{lipsum}		
\usepackage{graphicx}
\usepackage{doi}
\usepackage{authblk}
\usepackage{framed,multirow}

\usepackage{amssymb}
\usepackage{latexsym}
\usepackage{graphicx}
\usepackage{amsmath}
\usepackage{amssymb}
\usepackage{booktabs} 
\usepackage{tabularx}
\usepackage{array}
\usepackage{booktabs} 
\usepackage{tabularx, multirow}
\usepackage{tabularx, multirow, booktabs}
\usepackage{subcaption}
\usepackage{algorithm}
\usepackage{algorithmic}
\usepackage{authblk}
\usepackage{float}
\usepackage[nameinlink, noabbrev]{cleveref} 

\crefformat{figure}{#2{\color{red}Figure~#1}#3}
\crefformat{section}{#2{\color{red}Section~#1}#3}
\crefformat{equation}{#2{\color{red}Equation~#1}#3}
\crefformat{table}{#2{\color{red}Table~#1}#3}
\crefformat{algorithm}{#2{\color{red}Algorithm~#1}#3}

\usepackage{url}
\usepackage{xcolor}
\newcolumntype{C}[1]{>{\centering\arraybackslash}p{#1}}

\newcolumntype{H}{>{\centering\arraybackslash}X}

\usepackage{enumitem}
\newlist{boldenum}{enumerate}{1}
\setlist[boldenum]{label=\textbf{(\arabic*)}}

\definecolor{newcolor}{rgb}{.8,.349,.1}

\newcommand{\nonumfootnote}[1]{
    \begingroup
    \renewcommand\thefootnote{}
    \footnotetext{#1}
    \endgroup
}
 
\title{ Edge Detection Quantumized: A Novel Quantum Algorithm for Image Processing }

\author[1]{\large Syed Emad Uddin Shubha}
\author[1]{\large Mir Muzahedul Islam}
\author[1]{\large Tanvir Ahahmed Sadi} 
\author[1]{\large Md. Hasibul Hasan Miraz}
\author[1, *]{\large M.R.C. Mahdy}

\affil[1]{Department of Electrical and Computer Engineering, North South University, Bashundhara, Dhaka}

\hypersetup{
pdftitle={A template for the arxiv style},
pdfsubject={q-bio.NC, q-bio.QM},
pdfauthor={David S.~Hippocampus, Elias D.~Striatum},
pdfkeywords={First keyword, Second keyword, More},
}

\begin{document}

\maketitle

\begin{abstract}
Quantum image processing is a research field that explores the use of quantum computing and algorithms for image processing tasks such as image encoding and edge detection. Although classical edge detection algorithms perform reasonably well and are quite efficient, they become outright slower when it comes to large datasets with high-resolution images. Quantum computing promises to deliver a significant performance boost and breakthroughs in various sectors. Quantum Hadamard Edge Detection (QHED) algorithm, for example, works at constant time complexity, and thus detects edges much faster than any classical algorithm. However, the original QHED algorithm is designed for Quantum Probability Image Encoding (QPIE) and mainly works for binary images. This paper presents a novel protocol by combining the Flexible Representation of Quantum Images (FRQI) encoding and a modified QHED algorithm. An improved edge outline method has been proposed in this work resulting in a better object outline output and more accurate edge detection than the traditional QHED algorithm.
\end{abstract}


\section{Introduction}
\nonumfootnote{* Corresponding author.
\textit{E-mail address}: \href{mahdy.chowdhury@northsouth.edu}{mahdy.chowdhury@northsouth.edu} (M.R.C. Mahdy).}

In modern technology, image processing is an indispensable tool, employing sophisticated algorithms to extract valuable insights from visual data. Its applications are vast and diverse, spanning fields such as image segmentation, enhancement, computer vision, recognition, and medical imaging.  With the emergence of quantum image processing algorithms, significant strides have been made in critical operations like edge detection and beyond \cite{jing2022recent, wang2022review}.

Quantum computers have the prospect of revolutionizing many fields by solving problems that are currently computationally expensive.  The development of quantum computers is a rapidly advancing field of research and potential applications of quantum computers are currently being investigated. Quantum Image Processing shows a promising aptness by utilizing quantum superposition and entanglement as a powerful resource \cite{Alexander, cai2018survey, soklakov2006efficient}. Converting image data into a quantum state, we can perform a wide range of linear operations, \cite{Hans, kai, ruan2016quantum}, thus transforming and processing images for various applications. There are many Quantum Image Representation techniques, such as Flexible Representation of Quantum Images (FRQI), Quantum Probability Image Encoding (QPIE), and Novel Enhanced Quantum Representation (NEQR), offering efficient pathways for translating pixel information into quantum data.  Among these two methods, the FRQI method uses angle encoding, which requires fewer qubits \cite{le2009flexible, le2011flexible}. The QPIE method requires one qubit less than FRQI. The NEQR method, on the other hand, requires a much higher number of qubits. \cite{Tao, AhmedYounus}.  

Edge detection, a cornerstone of image analysis and feature extraction, assumes paramount importance across a multitude of applications, ranging from fingerprint recognition to medical imaging. While classical edge detection algorithms, such as, Sobel, Laplacian, Prewitt, and Canny, have proven effective in many scenarios, they encounter significant scalability challenges when dealing with larger images, owing to the increased computational demands \cite{Artyom,Chetia2021QuantumIE,Xu2020QuantumIP,Zhou2019QuantumIE}.

The emergence of quantum algorithms for edge detection presents a promising avenue for overcoming these challenges. Leveraging the inherent parallelism of quantum computation, these algorithms enable the simultaneous processing of multiple pixels, thereby reducing the overall time complexity. Among these pioneering algorithms, Quantum Hadamard Edge Detection (QHED) stands out for its constant time complexity and memory efficiency \cite{Gultom, yao2017quantum}. However, despite its inherent advantages, QHED's effectiveness diminishes when confronted with real-world images, exhibiting inaccuracies and noise due to its reliance on binary image bases.

In response to these challenges, this paper proposes a series of innovative modifications to enhance QHED's accuracy and efficiency. At first, we used the angle encoding (FRQI) method to convert a classical image into a quantum state, followed by a partial measurement. After that, we reset the existing qubit and use it as a resource to implement traditional QHED with an additional qubit method. Then a classical post processing has been applied to generate an edge-detected output image. We have also analyzed the procedure from an information-theoretic perspective. We have designed the protocol to encompass the development of a versatile quantum circuit adaptable to diverse image inputs, refinement of object outline methodologies to improve precision in representation, and significant enhancements in edge detection for complex images through a combination of classical and quantum modifications applied to existing QHED algorithms. Through these endeavors, we seek to push the boundaries of quantum image processing, paving the way for more effective and robust edge detection methodologies in both theory and practice. Fig. 1 shows our proposed protocol and Fig. 2 gives a comprehensive overview of our improved protocol over traditional QHED. 

\begin{figure*}[h]
    \centering
    {\includegraphics[width=0.9\linewidth]{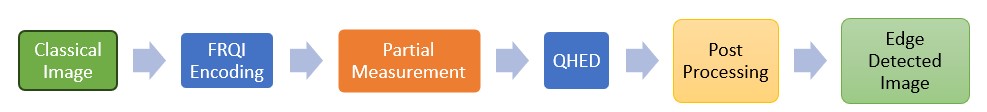}}
    \caption{System Diagram of the Proposed Model}
    \label{fig:process}
\end{figure*}

\begin{figure*}[h]
    \centering
    {\includegraphics[width=.9\linewidth, height = 30em]{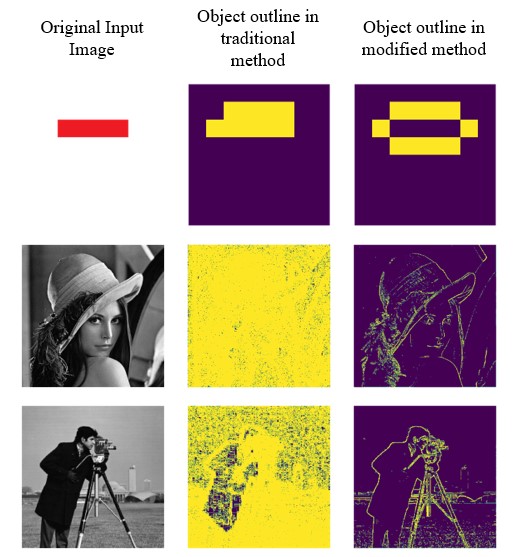}}
    \caption{Comparison between Traditional QHED and Proposed Model.}
    \label{fig:comparison}
\end{figure*}

\section{Preliminary Concepts}
\subsection{Quantum Image Processing}

An image is given by a matrix $F=(F_{ij})_{M \times N}$. If the image is a grey-scale image, then $0\leq M_{kl} \leq 1$ represents the pixel value of $j^{th}$ row and $k^{th}$ column. If the image is a color image, then we have $M_{lk}= [R_{lk} G_{lk} B_{lk}]$, representing the RGB values of a pixel, typically these values are 8 bits each, thus having a value between $0$ to $255$. Some images have an additional component, named channel.  
Classically image processing can be given by the Map $G=PFQ$, where $G=(G_{ij})_{M \times N}$. \\
In quantum image processing, We first need to encode the image into a quantum state called Quantum Image Representation. The state is represented by a normalized column vector $|f\rangle$ that encodes color and position information. There are several ways to encode an image, for example, in Novel Enhanced Quantum Representation (NEQR), a $2^n \times 2^n $ grey scale image is represented as \cite{zhang2013neqr, Engin} 
\begin{equation}
|I\rangle= \frac{1}{2^n}\sum_{Y=0}^{2^n-1} \sum_{X=0}^{2^n-1}|{f(Y,X)}\rangle|{YX}\rangle
\end{equation} 
Here, $\Omega_{YX} = \otimes_{i=0}^{i=q-1} |C_{YX}^{i} \rangle$ encodes pixel information, where $C_{YX}^i \in {0,1}$ and $q=8$ for grey scale image.  \\
Now quantum image processing can be described by linear quantum operations, given by:
\begin{equation}
\epsilon (|f\rangle)= \sum_{i}E_i|f\rangle \langle f| E_i^\dag
\end{equation} 
Here $\sum_{k} E_k^\dag E_k = \mathds{1}$. We call $\{E_j\}$ Kraus Operators, and the map $\epsilon$ represents a completely positive trace-preserving map, and $\epsilon (|f\rangle)$ is called a density operator.  

\subsection{Flexible Representation of Quantum Images}
The Flexible Representation of Quantum Images encodes the pixel data in angles, which requires only one qubit to store pixel information, rest of the qubits stores position information. 
Given $\{\theta_0, \theta_1, ..., \theta_{4^{n}-1}\}  \quad  (\theta_i \in [0,\pi/2])$, the FRQI state is given by,
\begin{equation}
|I(\theta)\rangle= \frac{1}{2^n}\sum_{i=0}^{4^n-1}(cos \theta_i|{0}\rangle+sin \theta_i|{1}\rangle)\otimes |{i}\rangle
\end{equation}

We can transform $|0\rangle^{\otimes2n+1}$ into FRQI state, $|I(\theta)\rangle$ by applying unitary operator $\mathcal{R}\mathcal{H}$, with $\mathcal{H}= I \otimes H^{\otimes 2n}$ and $\mathcal{R}= \prod_{i=0}^{4^n-1}R_i$. where $R_i$ is a unitary matrix named controlled rotation, given by
\begin{equation}
R_i= I \otimes \sum_{j=0, j \neq i}^{4^n-1} |{j}\rangle \langle {j}| + R_y(2\theta_i) \otimes  |{i}\rangle \langle {i}| = I^{\otimes 2n+1}+(R_y(2\theta_i)-I) \otimes  |{i}\rangle \langle {i}\mid
\end{equation}

We can implement the controlled rotation $R_i$'s using $C^{2n}R_y$ (controlled $R_y$) and $X$ gate. Here, 
\begin{equation}
C^{2n}R_y(2\theta_i)= I^{\otimes 2n+1}+(R_y(2\theta_i)-I) \otimes  |{4^n-1}\rangle \langle {4^n-1}\mid
\end{equation}

We can convert $C^{2n}R_y$ into $R_i$ by applying NOT gates on the left and right sides. For example, in case of $2\times 2$ grey image, we see, $R_3= C^2R_y$, and 
\begin{equation}
\begin{split}
 R_2 &= (I \otimes I \otimes X) C^2R_y (I \otimes I \otimes X) \\
R_1 &= (I \otimes X \otimes I) C^2R_y (I \otimes X \otimes I) \\
R_0 &= (I \otimes X \otimes X) C^2R_y (I \otimes X \otimes X) \\   
\end{split}
\end{equation}

Therefore controlled rotations $R_i$ can be implemented by
$C^{2n}R_y(2\theta_i)$ and NOT operations.

Basically, applying $I \otimes P_i $ in the both side of $C^{2n}R_y$ to make the conversion:
\begin{equation}
P_i |{4^n-1}\rangle = |{i}\rangle 
\end{equation}
Here $P_i$ is the tensor product of some $I$ and $X$, which is a hermitian. Fig. 1 refers to the circuit of the FRQI method for an arbitrary $2\times 2$ color image.
\\ 
Say, one colored pixel has the RGB values $(R_i, G_i, B_i)$, then we encode the information as follows: 
\begin{equation}
\theta_i=cos^{-1}(\frac{R_i}{256}+\frac{G_i}{256^2}+\frac{B_i}{256^3})
\end{equation}
The advantage is, given $\theta_i$, it is easier to get correspondent $R_i, G_i, B_i$ by expressing $256^3cos(\theta_i)$ as $256$ base number.  Fig. 3 refers to a circuit implementation of a $2\times 2$ image.
\begin{figure*}[h]
    \centering
    {\includegraphics[width=.9\linewidth, height = 20em]{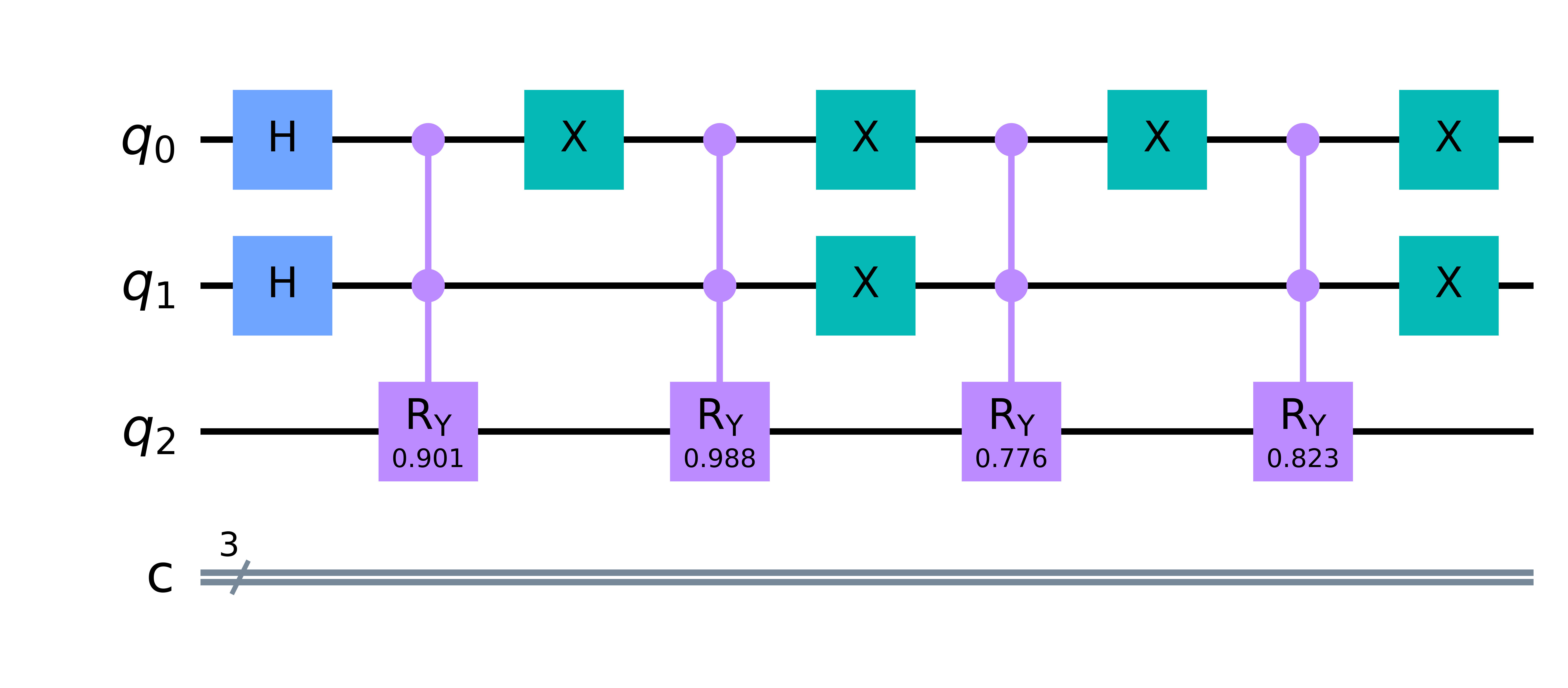}}
    \caption{FRQI Encoding circuit for a $2 \times 2$ sample color image.}
    \label{fig:FRQIl}
\end{figure*}

\subsection{Quantum Hadamard Edge Detection}
QHED edge detection algorithm is based on QPIE image representation. QPIE will convert digital image information into a quantum state.
To begin with, the images are represented as normalized probability amplitude corresponding to a specific quantum state. Given an image $I_{xy}$, it can be expressed as
\begin{equation}
|i m g\rangle=\sum_{x,y} \frac{I_{y x}}{\sqrt{\sum I_{y x}^2}}|xy\rangle = 
\begin{pmatrix}
c_0 & c_1 & \ldots & c_{N-1}
\end{pmatrix}
^{\mathrm{T}}
\end{equation}
To build the QHED circuit an additional qubit is required. This will help to calculate all the pixel differences at once.  At first, the image information is initialized to the N number of data qubits using the QPIE technique. Then, the Hadamard gate is applied to the auxiliary qubit. Finally, we have the following state that encodes the image: 
\begin{equation}
\left|i m g^{\prime}\right\rangle=\frac{1}{\sqrt{2}}
\begin{pmatrix}
c_0 & c_0 & c_1 & c_1 & \ldots & c_{N-1} & c_{N-1}
\end{pmatrix}
^{\mathrm{T}}
\end{equation}
After that, we performed an amplitude permutation unitary operation $D_{2^{n+1}}$ to shift down the element of the state vector, by one row. This operation will simultaneously facilitate the gradient calculation for both even and odd pixel pairs.
The expression of $D_{2^{n+1}}$ and its operation on the image is given in Eqn. (11) and (12) respectively.

\begin{equation}
D_{2^{n+1}}=\left[\begin{array}{ccccc}
0 & 1 & 0 & \cdots & 0 \\
0 & 0 & 1 & \cdots & 0 \\
\vdots & \vdots & \vdots & \ddots & \vdots \\
0 & 0 & 0 & 0 & 1 \\
1 & 0 & 0 & 0 & 0
\end{array}\right]
\end{equation}
\begin{equation}
D_{2^{n+1}}\left|i m g^{\prime}\right\rangle=\frac{1}{\sqrt{2}}
\begin{pmatrix}
c_0 & c_1 & c_1  & \ldots & c_{N-1} & c_{N-1} & c_0
\end{pmatrix}
^{\mathrm{T}}
\end{equation}
After the above operation, an H-gate is applied to the auxiliary qubit. The resultant state would be, 
\begin{equation}
\left|i m g^{\prime \prime}\right\rangle=\frac{1}{2}
\begin{pmatrix}
c_0+c_1 & c_0-c_1 & c_1+c_2 & c_1-c_2 \ldots c_{N-1}+c_0 & c_{N-1}-c_0
\end{pmatrix}
^{\mathrm{T}}
\end{equation}
The quantum circuit of the QHED algorithm for a $4\times 4$ binary image is shown in Fig. 4.
\begin{figure*}[h]
    \centering
    {\includegraphics[width=.9\linewidth, height = 15em]{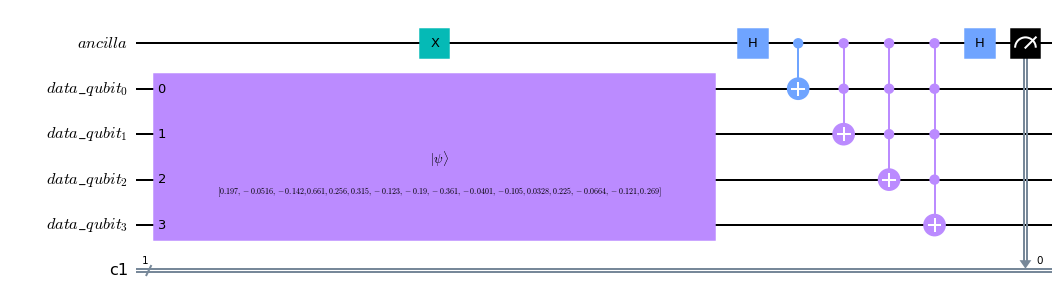}}
    \caption{Quantum Circuit of QHED Algorithm.}
    \label{fig:qiskitSim}
\end{figure*}
Edge detection is the process of detecting the edge of an object from a picture based on dissimilarity of color intensity. The discontinuation can be measured from the difference between two neighbor pixels. If we measured the auxiliary qubit being in state |1⟩ then the necessary information for edge detection can be retrieved. The whole procedure carries out a horizontal scan to detect edges in the horizontal direction. For vertical scan, the image is transposed and the above operations are repeated. Finally, the horizontal and vertical scan outputs are superimposed into one another using classical post-processing for the final edge detected output.  

\section{Proposed Edge Detection Algorithm}
In this paper, we have modified the edge outline of an object by analyzing the relative edge detection order and imposing a hyperparameter to filter out the noise. We have also modified the encoding part using FRQI and partial measurement. In this section, let's assume we have an image with size $2^n times 2^n$.
\subsection{Proposed Encoding Method}

Suppose we have two grey-scale images, Image$-1 = \{I_{i}\}$ and Image$-2 = \{I'_{i}\}$, such that $I'_{k}= \sqrt{1-I_{k}^2}$ ($ \forall{k}$). \\ Since $0\leq I_{k} \leq 1$, this map is a bijection with smooth behavior, that \textbf{preserves the edge} information, as shown in Fig. 5:

\begin{figure*}[h]
    \centering
    {\includegraphics[width=.6\linewidth, height= 15em]{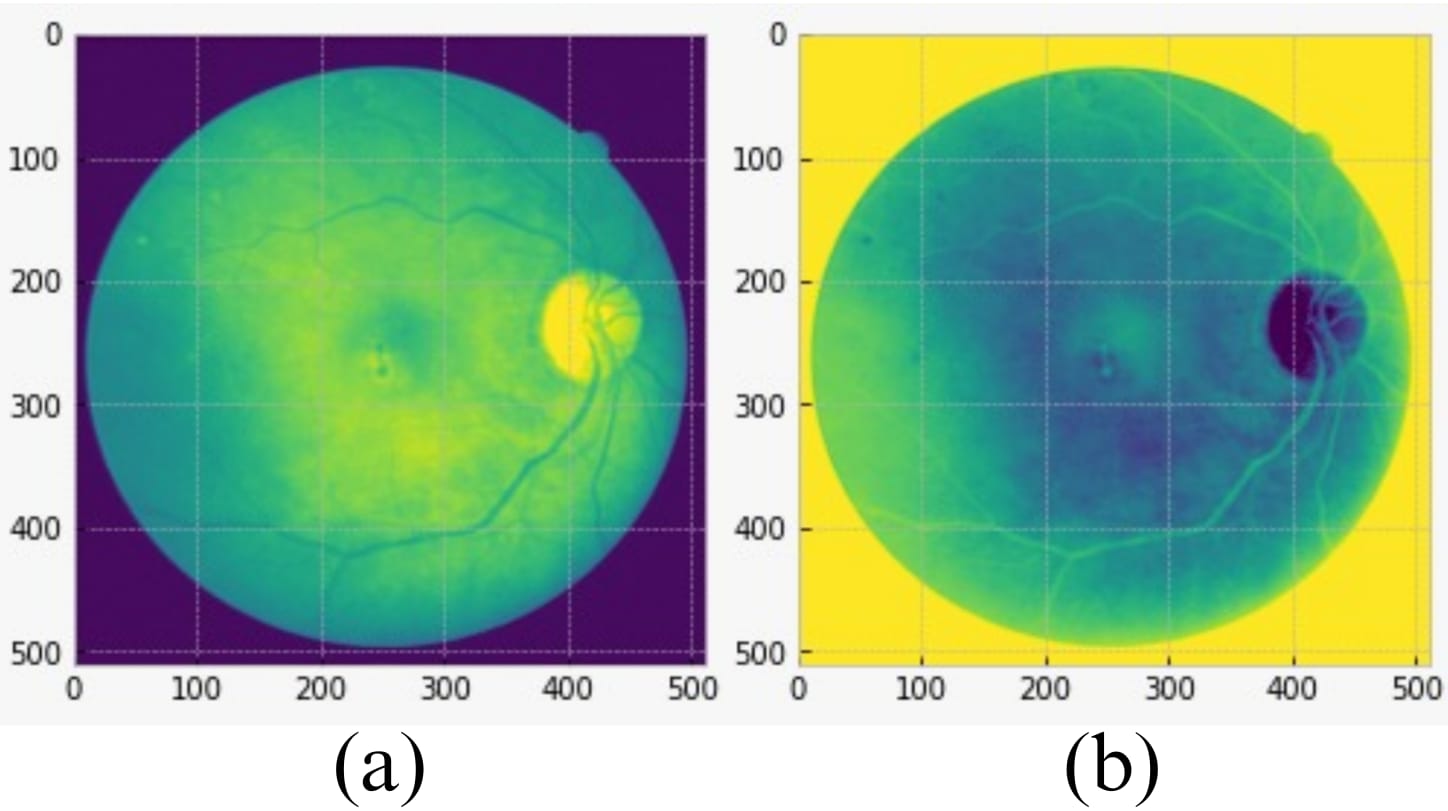}}
    \caption{(a) image $-1$, (b) image $-2$. We note that the map doesn't change the edges.}
    \label{fig:edge_thake}
\end{figure*}
Now, say, $I_i = cos(\theta _{i})$. Therefore the QPIE representations of Images 1 and 2 are given by, 
\begin{equation}
|img_1\rangle=\sum_{i=0}^{4^n-1} \frac{I_{i}}{\sqrt{\sum_{j=0}^{4^n-1} I_{j}^2}}|i\rangle = \sum_{i=0}^{4^n-1} \frac{cos(\theta_{i})}{\sqrt{\sum_{j}{ cos^2(\theta_j})}}|i\rangle
\end{equation}

\begin{equation}
|\text{img}_2\rangle = \sum_{i=0}^{4^n-1} \frac{I_i'}{\sqrt{\sum_{j=0}^{4^n-1} I_{j}'^2}} |i\rangle = \sum_{i=0}^{4^n-1} \frac{\sin(\theta_i)}{\sqrt{\sum_{j} \sin^2(\theta_j)}} |i\rangle 
\end{equation}
Now in our proposed protocol, we first encode our image using FRQI, given by Eq. (3). Say, the pixel information is stored at ancilla qubit, which will now be measured in the Z-basis, i.e., the Krauss Operators are given by, $\{|0 \rangle \langle 0| \otimes \mathds{1}^{\otimes 2n}, |1 \rangle \langle 1| \otimes \mathds{1}^{\otimes 2n} \}$. The collapsed state will be given by either Eq. (14) or Eq. (15), depending on the measurement outcome. Since both collapsed state resembles the same edge, we will now apply the QHED algorithm with the additional qubit. The modified algorithm is shown in the diagram below:

\begin{figure*}[h]
    \centering
    {\includegraphics[width=.9\linewidth]{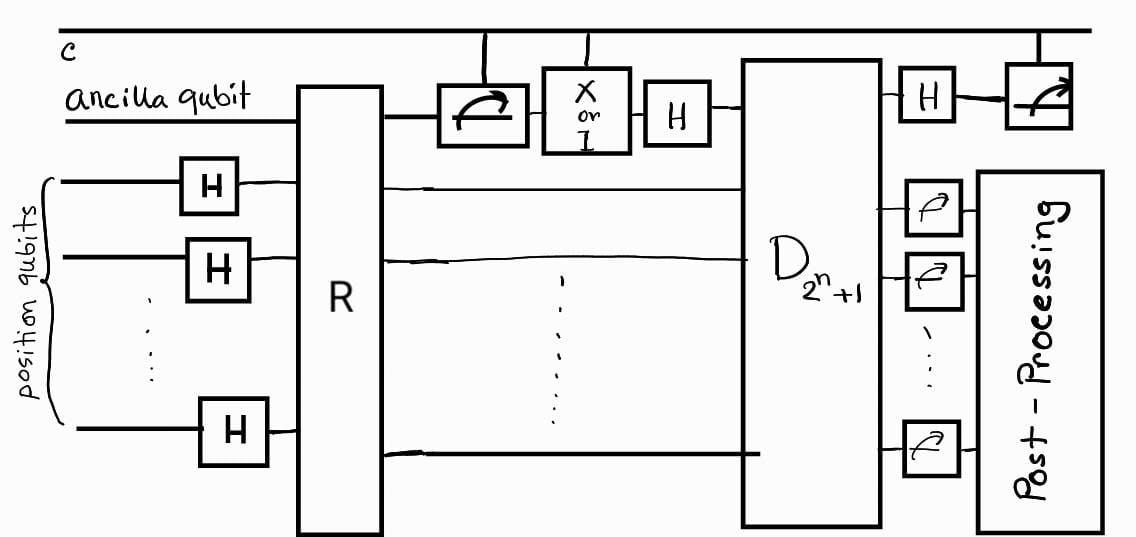}}
    \caption{System Diagram of Modified Algorithm.}
    \label{fig:modifiedQHED}
\end{figure*}
\subsection{Edge Detection Improvement}
The ancilla qubit will be measured at state $|0\rangle$ or $|1\rangle$, we will put an $X$ gate if the first case happens. After that a Hadamard gate will be applied to the ancilla qubit, that is how the state of the ancilla qubit will now be in $|-\rangle$ to apply conventional QHED. QHED process compares two neighbor pixels, and based on their dissimilarity, it plots the edge. The edge detection operation occurs based on the superimposed scan results of left-to-right and top-to-bottom scans. One of the issues with QHED edge detection is that, it does not consider object boundaries. If there is a difference in color intensity between the current and the next pixel, QHED portrays an edge in the current pixel. QHED defines the edges inside an object boundary and does not wrap around the object. As a consequence, the resultant object outline is ambiguous and does not represent the actual object and shape correctly.  We considered the object's exact position to address this complication and analyzed the pixel gradient result. The edge outline around the object borders are then portrayed according to these findings.

 Comparison of edge detection between traditional approach (middle column) and modified approach (far right column) for input image (far left column). It is evident the conventional approach does not capture the object outline correctly, and the actual object retrieval might not be easy. In comparison, our proposed model outlines the object accurately. The main QHED algorithm cannot detect all the edges properly for complex images and produces a lot of noise, while our modified approach can reduce noise using a dynamic threshold. In Fig. 7,8 \& 9, we have demonstrated how our edge detection algorithm provides better visualization. 
\begin{figure*}[h]
    \centering
    {\includegraphics[width=.9\linewidth, height = 30em]{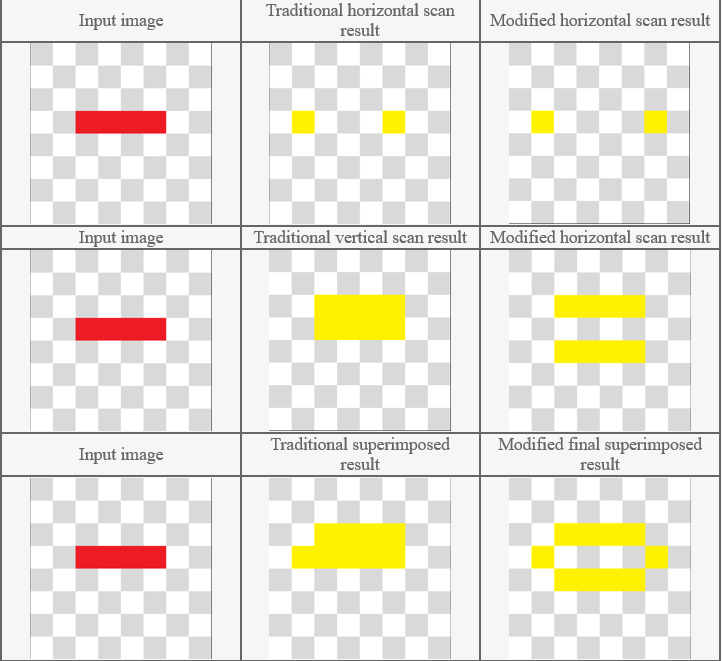}}
    \caption{Scanning result difference between traditional QHED and the proposed method.}
    \label{fig:qiskitSim}
\end{figure*}
\begin{figure}[H]
    \centering
    {\includegraphics[width=.9\linewidth, height = 30em]{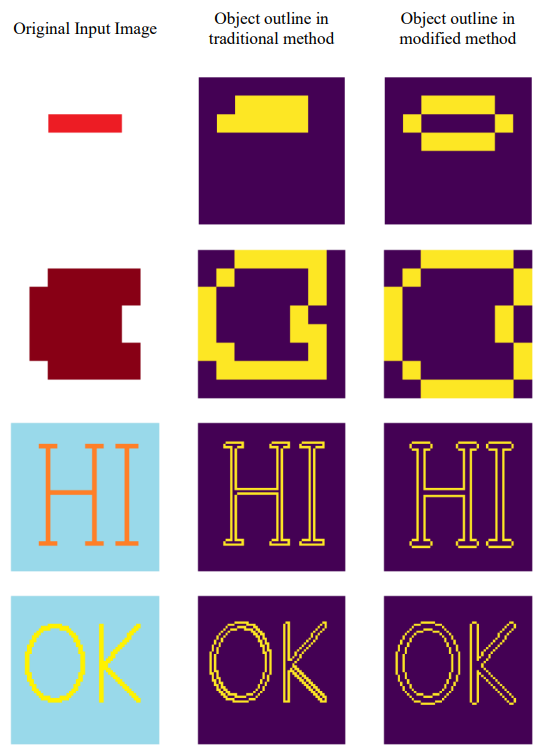}}
    \caption{Comparison of edge detection between traditional approach (middle column) and modified approach (far right column) for input image (far left column). It is evident the traditional approach does not capture the object outline correctly and the actual object retrieval might not be easy. Whereas our proposed model outlines the object accurately.}
    \label{fig:qiskitSim}
\end{figure}

\begin{figure}[H]
    \centering
    {\includegraphics[width=.9\linewidth, height = 60em]{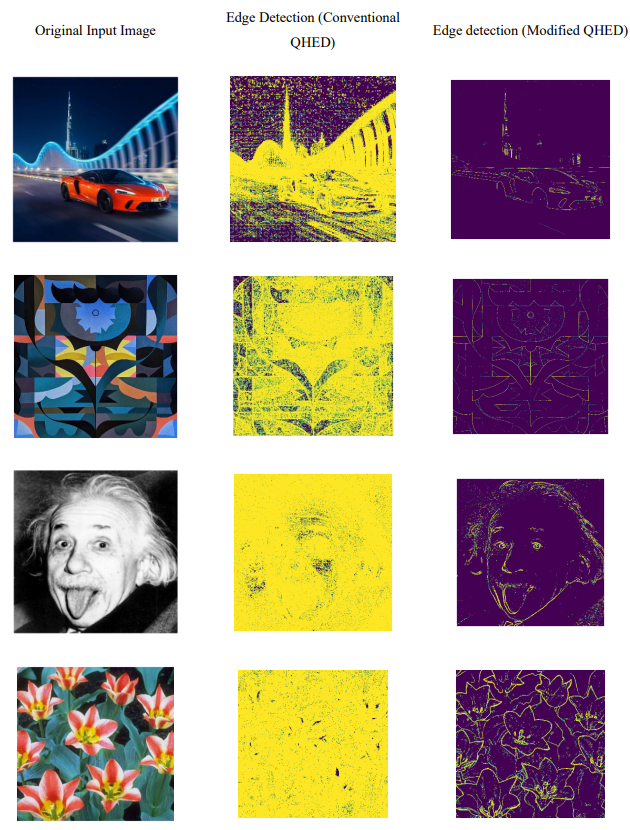}}
    \caption{Comparison of edge detection output. The middle column represents the conventional QHED implementation result, and the improved implementation of QHED is shown in the far-right column.}
    \label{fig:qiskitSim}
\end{figure}
Threshold has been used to remove noise in grey-scale images and improve the edge outline. For any $2^n \times 2^n$ image, if the represented state is given by $|\psi \rangle = \sum_{i} c_i|i\rangle$, we define the threshold as:
\begin{equation}
thr = \frac{|max_{i}\{ c_i-c_{i+1} \}|}{2n} 
\end{equation}
In the post-processing, we will generate a binary image, if the difference is greater than the threshold of the main picture, we will call it an edge. If the sign of difference is mismatched with the first edge, the edge will be shifted one pizel either in the vertical or horizontal direction based on scanning order. Lastly, noises will be removed if the pixel difference is less than $(-thr)$. Thus the post-processing gives us an improved edge outline without noise.
\subsection{Quantum Thermodynamics Perspective}
When we partially measure the state given by Eq. (3), we have either collapsed the state in Eq.(14) or Eq. (15). We note that, even if we were to use Eq. (8) to preserve the color information, the measurement will make the state into the representation of a grey-scale image. We can easily note that by observing the fact that, for $i=1,2,..,n$, the set $\{cos(\theta_i)\}$ is independent but $\{\frac{cos(\theta_i)}{\sum_i {cos^2(\theta_i}}\}$ is not. Hence, it can be said that grey-scale representation loses some information, we can produce the state by partially measuring the color-encoded state, i.e., measuring the pixel qubit. The Edge detection overall means losing some information but keeping the edge outline invariant. This also will be examined thoroughly in the next version of this work.

\section{Conclusion}
In this paper, we provided a generalized implementation scheme for NEQR and FRQI algorithms. We also have addressed the shortcomings of the QHED algorithm and proposed a modified version of QHED. Traditional QHED does not work for all input images, and it struggles with complex images. Our modified algorithm allowed the encoding and edge-detecting circuits to work with various types of complex images. We implemented a dynamic circuit for the input image, fixed the edge outline, and, most importantly, adjusted the threshold value to remove unnecessary noise. Overall, our project has significantly improved the QHED algorithm, making it a more efficient and accurate method for edge detection in quantum image processing. The changes we have made will allow the QHED algorithm to be applied to a broader range of images and provide more accurate results. We have also used the FRQI method, along with a partial measurement, to replace the QPIE method. This inclusion makes our model a more novel and physics-oriented approach. 

\section*{Acknowledgment} 
The authors would like to thank Research Assistants Deponker Sarkar Depto and Md Shawmoon from the NSU Optics Lab for their constructive suggestions and technical help, respectively. M.R.C. Mahdy acknowledges the support of NSU CTRGC grant 2023-24 (approved by the members of BOT) and NSU internal grant.


\end{document}